\documentclass[11pt, a4paper]{article}
\usepackage{jcappub}

\usepackage[utf8]{inputenc} 
\usepackage[T1]{fontenc}    
\usepackage{lmodern}
\usepackage{hyperref}       
\usepackage{url}            
\usepackage{amsfonts}       
\usepackage{amsmath,amsthm,amssymb} 
\usepackage{bm}
\usepackage{bbm}
\usepackage{dcolumn}
\usepackage{nicefrac}       
\usepackage{graphicx}
\usepackage{xcolor, etoolbox}
\usepackage{mathtools}
\usepackage{mathrsfs}
\usepackage{soul}
\usepackage{ragged2e}
\usepackage{microtype}      
\usepackage{braket}      
\usepackage{enumitem}
\usepackage{xspace}
\usepackage{dsfont}
\usepackage{float}
\usepackage[export]{adjustbox}
\usepackage{lipsum} 
\usepackage[capitalise]{cleveref} 
\crefrangelabelformat{equation}{(#3#1#4--#5#2#6)}
\crefname{equation}{Eq.}{Eqs.}
\Crefname{equation}{Equation}{Equations}
\usepackage{overpic}
\usepackage[normalem]{ulem}

\usepackage[normalem]{ulem} 

\DeclarePairedDelimiter\abs{\lvert}{\rvert}

\renewcommand{\vec}[1]{\boldsymbol{#1}}
\newcommand{\dd}{{\text{d}}}
\newcommand{\diff}{{\text{d}}}

\newcommand{\etaf}{{\eta_{\text{f}}}}

\newcommand{\mpl}{{M_{\text{P}}}}

\newcommand{\vx}{{\bm{x}}}

\hypersetup{
    colorlinks,
    linkcolor=blue,
    citecolor=blue,
    urlcolor=blue,
    breaklinks=true 
}

\begin{document}


\title{\boldmath Influence of spatial curvature in cosmological particle production}


\author[a]{Jose A. R. Cembranos,}
\author[a]{Luis J. Garay,}
\author[a,b,c]{Álvaro Parra-López}
	
\affiliation[a]{Departamento de F\'isica Te\'orica and IPARCOS,
Universidad Complutense de Madrid, \\ 28040 Madrid, Spain} 
\affiliation[b]{Instituto de F\'isica Te\'orica UAM/CSIC, calle Nicol\'as Cabrera 13-15,\\ Cantoblanco, 28049, Madrid, Spain}
\affiliation[c]{Department of Physics, University of Oslo, N-0316 Oslo, Norway}

\emailAdd{cembra@fis.ucm.es}
\emailAdd{luisj.garay@ucm.es}
\emailAdd{alvaparr@ucm.es}
	
\date{\today}

\abstract{We analyze cosmological particle production driven by spacetime expansion in the early universe for homogeneous and isotropic cosmologies with positive, negative, and zero spatial curvature. We prioritize analytical results to gain a deeper understanding of curvature-induced effects. Specifically, for a conformally coupled scalar field, we model the inflationary epoch as an exact de Sitter phase followed by a transition to a static universe. Both instantaneous and smooth exits from inflation are considered, the latter being implemented via the adiabatic vacuum prescription. Starting from an initial Bunch-Davies vacuum, we derive the associated mode functions carefully adapted to each curvature sign. Using the Bogoliubov formalism, we non-perturbatively compute the number density of produced scalar particles. Our results demonstrate that spatial curvature significantly impacts the resulting particle spectra, particularly for light fields, where the deviation from the flat-space scenario is most prominent and can reach several orders of magnitude.}


\arxivnumber{-}

\maketitle
\flushbottom


\section{Introduction}
\label{sec:introduction}

Cosmological particle production due to the expansion of spacetime is a well-established prediction of quantum field theory in curved spacetimes \cite{Birrell1982,Mukhanov2007,Parker2009}, originally studied by Parker in the context of the early universe~\cite{Parker1968,Parker1969,Parker1971}. This mechanism has important implications for the generation of primordial perturbations \cite{Bardeen1980,Mukhanov1988,Liddle2000} during inflation \cite{Guth1981,Linde1982,Albrecht1982}, but is also capable of explaining the production of spectator dark matter with negligible interactions with the Standard Model~\cite{Chung1998a,Chung1998b,Kolb1998,Chung2001,Graham2015,Ema2016,Markkanen2017,Markkanen2018,Ema2018,Ema2019,Bastero2019,Fairbairn2019,Wang2019,Chung2019,Hashiba2019,Tenkanen2019,Herring2020,Borrajo2020,Ahmed2020,Cembranos2020,Sato2022,Lebedev2022,Kainulainen2023,Garcia2023a,Cembranos2023,Cembranos2024,Capanelli2024a,Capanelli2024b,Ozsoy2024,Garcia2024b,Racco2024,Garcia2025,Verner2025,deHaro2025}.

Most studies of cosmological particle production assume a spatially flat Friedmann-Lemaître-Robertson-Walker (FLRW) universe \cite{Friedman1922,Friedman1924,Lemaitre1931,Robertson1935,Robertson1936a,Robertson1936b,Walker1937}. However, current observational constraints still allow for a small but nonzero spatial curvature~\cite{Planck2018-1}, which may affect the particle production process. Furthermore, from a theoretical perspective, the spatial curvature could have been significantly large during the pre-inflationary era, potentially playing a dominant role in shaping the initial states of cosmic evolution before the exponential expansion drove the universe toward flatness. Indeed, some studies have shown that spatial curvature modifies the behavior of low wavenumbers in the cosmological power spectrum, for which the curvature is most relevant \cite{Bonga2016,Bonga2017,Handley2019,Hergt2022}.

When studying particle production in curved spacetimes, one must address the ambiguity in the choice of vacuum state---inherent to quantum field theories \cite{Alvarez2023}---by selecting specific quantizations. In cosmological contexts, the most suitable choices are typically related to the characteristics of the spacetime geometry. In particular, the natural choice in FLRW universes is the Bunch-Davies vacuum. However, different spatial curvatures lead to distinct evolution of the field modes, which in turn modifies the predicted density of produced particles.

In this work, we study cosmological particle production in the early universe for a scalar field conformally coupled to the geometry, in the context of FLRW universes with closed, flat, and open spatial sections. We consider an inflationary model based on an exact de Sitter phase followed by a transition to a radiation-dominated universe, which is considered to be static regarding particle production. We model this transition either instantaneously, or by implementing a smooth transition to staticity---via the adiabatic vacuum prescription~\cite{Parker1974}. These approximations are adopted to ensure the tractability of the model, allowing us to obtain exact analytical results and, consequently, to gain better control and a deeper understanding of the specific role that spatial curvature plays in the process. 

The different spatial curvature cases can be understood as different slices of the full de Sitter spacetime. For massive scalar fields in de Sitter spacetime, there is a preferred choice of vacuum, the so-called Bunch-Davies state \cite{Bunch1978}. To compute production, we introduce the Bunch-Davies mode functions adapted to each curvature case, and compute the Bogoliubov 
coefficients relating this notion of vacuum with the one natural to an observer living after the expansion. This allows us to obtain the spectra of produced particles for both instantaneous and smooth transitions to staticity. Lastly, we discuss the implications for the cosmological production of spectator fields and assess whether spatial curvature can significantly affect present-day abundances.

The structure of this paper is as follows. In Section~\ref{sec:scalarfield}, we review the dynamics of a scalar field in general FLRW cosmologies with arbitrary spatial curvature. Section~\ref{sec:deSitter} introduces the de Sitter model as a simple realization of inflation, including the transition towards radiation domination. In Section~\ref{sec:desitter}, we detail the quantization procedure in de Sitter spacetime and the choice of vacuum, for the three different spatial curvature cases. Section~\ref{sec:particleproduction} presents our results on particle production and abundances, while we elaborate our conclusions in Section~\ref{sec:conclusions}.

\textit{Notation}. We set $\mpl = 1/\sqrt{G},\, \hbar=c=k_B=1$, and use  the signature $(-,+,+,+)$. Greek indices run from $0$ to $3$.

\section{Dynamics of a scalar field in general FLRW cosmologies}
\label{sec:scalarfield}

The dynamics of a massive scalar field $\varphi$ non-minimally coupled to gravity in a $(1+3)$-dimensional Friedmann-Lemaître-Robertson-Walker (FLRW) universe   is given by the action 
\begin{equation}
S = -\frac{1}{2}\int\text{d}^4x \sqrt{-g}\left[g^{\mu\nu}\partial_{\mu}\varphi \partial_{\nu}\varphi + \left(m^2 + \xi R\right)\varphi^2\right],
\label{eq:GeneralAction}
\end{equation}
where $m$ is the bare mass of the field, $\xi$ denotes the coupling strength to the curvature scalar $R$, and $g$ is the determinant of the metric. The latter is characterized by the line element
\begin{equation}
\text{d}s^2 = a^2(\eta)\left[-\text{d}\eta^2 +
 \frac{\text{d}r^2}{1-\kappa r^2} + r^2\dd \Omega^2(\theta, \phi)\right],
 \label{eq:GeneralLineElement}
\end{equation}
where the relation between conformal time $\eta$ and cosmological time $t$ is given by $a(\eta)\dd \eta = \dd t$, and $\kappa$ is the curvature of the spatial sections, which can be positive, negative or zero. Here, $\dd \Omega^2$ denotes the line element of the unit 2-sphere, with $\theta$ and $\varphi$ being the polar and azimuthal angles. Note that the universe has a finite spatial size $r_{\text{max}} = 1/\sqrt{\kappa}$ in the case $\kappa > 0$, whereas it can be assumed to be infinite otherwise. It will be convenient to perform the change of coordinates  
\begin{equation}
    r = \begin{cases}
    \abs{\kappa}^{-1/2}\sin{\left(\sqrt{\abs{\kappa}}u\right)} \quad &\text{with} \quad u \in [0, \pi/\sqrt{\abs{\kappa}}), \quad \text{for} \quad \kappa>0, \\
    u \quad &\text{with} \quad u \in [0, \infty), \quad \hspace{0.83cm} \text{for} \quad \kappa=0, \\
    \abs{\kappa}^{-1/2}\sinh{\left(\sqrt{\abs{\kappa}}u\right)} \quad &\text{with} \quad u \in [0, \infty), \quad \hspace{0.83cm} \text{for} \quad \kappa<0,\\
    \end{cases}
\label{eq:RadialCoordTransf}
\end{equation}
so that the line element can be written in compact form as
\begin{equation}
\text{d}s^2 = a^2(\eta)\left\{-\text{d}\eta^2 +
 \left[\text{d}u^2 + r ^2(u)\text{d}\Omega^2(\theta,\phi)\right]\right\},
\end{equation}

It is customary to work with the rescaled field $\chi(\eta, \vec{u}) = a(\eta)\varphi(\eta, \vec{u})$, whose equation of motion in coordinates $\bm{u} = (u, \theta, \phi)$ is given by,
\begin{equation}
\chi^{\prime\prime}(\eta, \vec{u}) - \left\{\Delta - \kappa - a^2(\eta)\left[m^2 + \left(\xi - 1/6\right) R\right]\right\}\chi(\eta, \vec{u})=0,
\label{eq:EOM}
\end{equation}
where the prime denotes derivative with respect to conformal time, we used the expression for the scalar curvature \mbox{$R = 6\left(a^{\prime\prime}/a^3 + \kappa/a^2\right)$}, and $\Delta$ is the Laplace-Beltrami operator of the spatial part of the metric,
\begin{equation}
    \Delta = \frac{1}{r^2(u)}\left[\partial_{u}\left(r^2(u)\partial_{u}\right) + \Delta_{\Omega}\right],
\label{eq:LaplaceBeltrami}
\end{equation}
where $\Delta_{\Omega}$ is the Laplace operator on the unit two-sphere.

The field $\chi$ can be expanded in terms of the eigenfunctions of this Laplacian, which are of the form \mbox{$\mathcal{H}_{klq} (u, \theta, \phi)=\mathcal{U}_{kl}(u)Y_{lq}(\theta, \phi)$}, where $Y_{lq}$ are the spherical harmonics  and the explicit form of $\mathcal{U}_{kl}(u)$ together with the details are given in \cref{sec:eigenfunctions3Dcurved}. The corresponding eigenvalues read
\begin{equation}
    \lambda^2(k) = \begin{cases}
    {\bar k}({\bar k}+2)\abs{\kappa} \quad &\text{for} \quad \kappa>0, \\
    k^2 \quad &\text{for} \quad \kappa = 0, \\
    ({\bar k}^2 + 1)\abs{\kappa} \quad &\text{for} \quad \kappa<0,
\label{eq:Eigenvalues}
\end{cases}
\end{equation}
with ${\bar k} = k/\sqrt{\abs{\kappa}}$ and
\begin{align}
    {\bar k} &= 0, 1, \text{...}, \infty; \quad l = 0, 1, \text{...}, {\bar k}, \quad &&\text{for} \quad \kappa > 0, \label{eq:LabelsDescriptionA} \\ 
    k &\in [0, \infty); \hspace{0.65cm}\quad l = 0, 1, \text{...}, \infty, \quad &&\text{for} \quad \kappa \leq 0, \label{eq:LabelsDescriptionB} \\ 
    q &= -l, -l+1, \text{...}, 0, \text{...}, l-1, l, \quad &&\text{for all values of} \kappa. \label{eq:LabelsDescriptionC}
\end{align}
If one takes the limit $\kappa \to 0$ in the eigenfunctions for the curved cases one recovers that of the flat geometry (details in \cref{sec:eigenfunctions3Dcurved}). Interestingly, the smallest eigenvalue in the negatively curved case is non-vanishing, acting as an effective mass~gap. 

With this, we can expand the field $\chi$ as
\begin{equation}
\chi(\eta,\vec{u}) = \int \dd \mu_{klq} \left[a_{klq}\mathcal{H}_{klq}(\vec{u})v_k(\eta)
+ a_{klq}^*\mathcal{H}_{klq}^*(\vec{u})v_k^*(\eta)\right],
\label{eq:FieldExpansion}
\end{equation}
where the measure in momentum space is given by
\begin{equation}
\int \dd \mu_{klq} = \int\diff\mu_k \sum_{l=0}^{K} \sum_{q=-l}^l, 
\end{equation}
where $K=\infty $ for $\kappa\leq 0$, $K=\bar k$ for positive $\kappa$, and
\begin{equation}
\int \dd \mu_{k} = \begin{cases}
\displaystyle\sum_{{\bar k}=0}^{\infty}\abs{\kappa}^{3/2}({\bar k}+1)^2  \quad &\text{for} \quad \kappa>0,\\
\displaystyle\int_0^{\infty}\text{d}k \, k^2 \quad &\text{for} \quad \kappa \leq 0.
\label{eq:PartialMomentumMeasure}
\end{cases}
\end{equation}

Introducing the expansion \eqref{eq:FieldExpansion} in the equation of motion \eqref{eq:EOM}, one finds that the time-dependent mode functions $v_k(\eta)$ satisfy a decoupled system of ordinary differential equations of the form
\begin{equation}
v^{\prime\prime}_k(\eta) + \omega_k^2(\eta) v_k(\eta) = 0.
\label{eq:ModeEquation}
\end{equation}
These are harmonic oscillator equations with time-dependent frequencies, given by
\begin{equation}
\omega_k^2(\eta) = \lambda^2(k) + a^2(\eta)\left[m^2+ \left(\xi - 1/6\right)R(\eta)\right].
\label{eq:Frequency}
\end{equation}
Solutions to \cref{eq:ModeEquation} are normalized with the Wronskian as
\begin{equation}
\text{Wr}\left[v_k, v_k^*\right] =v_kv^{\prime\, *}_k - v^{\prime}_k v_k^* = i,
\label{eq:WronskianModes}
\end{equation}
so that they are compatible with the standard Klein-Gordon product associated with differential equation of motion of the field \cite{Birrell1982,Mukhanov2007}. 

The modes~$v_k$ and $v^*_k$ constitute one basis of the space of solutions of the mode equation~\eqref{eq:ModeEquation}, and lead to an expansion in terms of the coefficients $a_{klq}$ and $a^*_{klq}$, as we have written in \cref{eq:FieldExpansion}. Nevertheless, one may choose a different basis for expanding the field, say, some other modes $u_k$ and $u^*_k$. The corresponding expansion coefficients will be, of course, different, $b_{klq}$ and $b^*_{klq}$. Writting the $u_k$ modes in terms of the $v_k$ modes allows one to find a relation between the two sets of solutions to the mode equation, 
\begin{equation}
    u_k = \alpha_k v_k + \beta_k v_k^*,
\label{eq:BogoliubovTransformationModes}
\end{equation}
where~$\alpha_k$ and~$\beta_k$ are the so-called Bogoliubov coefficients, and fulfill the relation $\abs{\alpha_k}^2 - \abs{\beta_k}^2 = 1$ (enforced by \cref{eq:WronskianModes}). These coefficients can be written in terms of the Wronskian of the modes as 
\begin{equation}
    \alpha_k = -i\text{Wr}[u_k, v_k^*], \quad \beta_k = i\text{Wr}[u_k, v_k].
\label{eq:BogoliubovCoefficientsWronskian}
\end{equation}
Since the field can be expanded in either basis, one must have that
\begin{equation}
\begin{split}
    \chi(\eta, \vx) =& \int \dd \mu_{klq} \left[a_{klq}  \mathcal{H}_{klq}(\vx)  v_k(\eta) + a_{klq}^*  \mathcal{H}_{klq}^*(\vx)  v_k^*(\eta)\right] \\
    =& \int \dd \mu_{klq}\left[b_{klq} \mathcal{H}_{klq}(\vx) u_k(\eta) + b_{klq}^* \mathcal{H}_{klq}^*(\vx) u_k^*(\eta)\right].
\end{split}
\end{equation}
Introducing the Bogoliubov transformation between the modes \eqref{eq:BogoliubovTransformationModes} in the expression above, one can find the relation between the coefficients $a_{klq}$ and $b_{klq}$, which reads (note that $\mathcal{H}_{klq}^* = (-1)^q \mathcal{H}_{kl,-q}$)
\begin{equation}
    b_{klq} = \alpha_k^* a_{klq} - (-1)^q\beta_k^*a_{kl,-q}^* \quad \text{and} \quad a_{klq} = \alpha_k b_{klq} + (-1)^q\beta_k^*b_{kl,-q}^*.
\label{eq:BogoliubovTransformationCoefficients}
\end{equation}

We will quantize the field $\chi$ following the canonical approach, to which we dedicate the following subsection. 

\subsection{Canonical quantization}

Canonical quantization implies promoting the expansion coefficients $a_{klq}, a_{klq}^*$ to creation and annihilation operators (and likewise for the $b_{klq}$, $b_{klq}^*$ coefficients), which fulfill the commutation relations
\begin{equation}
\big[\hat{a}_{klq}^{\dagger},\hat{a}_{k^{\prime}l^{\prime}q^{\prime}}^{\dagger}\big] = \left[\hat{a}_{klq},\hat{a}_{k^{\prime}l^{\prime}q^{\prime}}\right]=0, \quad
    \big[\hat{a}_{klq}, \hat{a}_{k^{\prime}l^{\prime}q^{\prime}}^{\dagger}\big] = \delta^\mu(klq,k'l'q'),
\label{eq:CreationAnnihilationCommutationRelsScalar}
\end{equation}
where
\begin{equation}
   \delta^\mu(klq,k'l'q')= \delta_{ll^{\prime}} \delta_{qq^{\prime}}\times \begin{cases}
    \displaystyle\frac{\delta_{{\bar k} {\bar k}^{\prime}}}{\abs{\kappa}^{3/2}{\bar k}^2} \quad &\text{for} \quad \kappa > 0,\\ ~\\
    \displaystyle\frac{\delta(k - k^{\prime})}{k^2} \quad &\text{for} \quad \kappa \leq 0\\
    \end{cases}
\end{equation}
and satifies $\int\diff\mu_{klq}\delta^\mu(klq,k'l'q')=1$. These are consistent with the commutation relations
\begin{equation}
    \left[\hat{\chi}(t, \vec{u}), \hat{\pi}_{\chi}(t, \vec{u}^{\prime}) \right] =i\delta(\theta - \theta^{\prime})\delta(\phi - \phi^{\prime})\delta(u - u^{\prime})
\label{eq:FieldsCommutationRelsScalar}
\end{equation}
on the field $\hat{\chi}$ and its conjugate momentum $\hat{\pi}_{\chi}$. 

These operators define the vacuum state $\ket{0^a}$:
\begin{equation}
\hat{a}_{klq}\ket{0^a} = 0, \quad \forall\, {klq}.
\end{equation}
By acting with the creation operator on the vacuum state, one can build a basis of the corresponding Hilbert space (the so-called Fock space~\cite{Birrell1982}). Note that, in general, each set of coefficients, $a_{klq}$ and $b_{klq}$ (and their complex conjugates), associated with the basis $v_k$ and $u_k$, respectively, defines two different notions of vacuum $\ket{0^a}$ and $\ket{0^b}$, respectively \cite{Mukhanov2007}. Then, the Bogoliubov transformation between the coefficients \eqref{eq:BogoliubovTransformationCoefficients} becomes a relation between the two sets of creation and annihilation operators,
\begin{equation}
    \hat{b}_{klq} = \alpha_k^* \hat{a}_{klq} - (-1)^q\beta_k^*\hat{a}_{kl,-q}^{\dagger} \quad \text{and} \quad \hat{a}_{klq} = \alpha_k \hat{b}_{klq} + (-1)^q\beta_k^*\hat{b}_{kl,-q}^{\dagger}.
\label{eq:BogoliubovTransformationOperators}
\end{equation}

The number density of $b$-particles in the $a$-vacuum, which will be, in general, a non-vacuum state according to the $\hat{b}_{klq}$ operators, turns out to be independent of $q$ and $l$ and is given by
\begin{equation}
    n_k^b = \abs{\beta_k}^2/(2\pi^2).
\label{eq:MeanNumberDensity}
\end{equation}
Integrating over all modes, we find the total mean density $n = \int \dd \mu_{klq} n_k$. As long as the latter remains finite, the Bogoliubov transformation~\eqref{eq:BogoliubovTransformationOperators} is well-defined, and the dynamics is implemented unitarily~\cite{Alvarez2022}. Let us see that this is the case. 

The total number density of particles is given by
\begin{align}
    n=\frac{1}{V}\int\diff\mu_{kql}\bra{0^a}b^\dagger_{klq}b_{klq}\ket{0^a}=\frac{1}{V}\int\diff\mu_{klq}\delta^\mu(klq,klq)|\beta_k|^2.
\end{align}
The factor $\delta^\mu(klq,klq)$ can be written using the completeness relation for the eigenfunctions~$\mathcal H_{klq}$ as
\begin{equation}
    \delta^\mu(klq,klq)=\int_0^\infty \diff u r^2(u)\diff\Omega \mathcal H_{klq}(\vec u)\mathcal H_{klq}^*(\vec u)=\int_0^\infty\diff u r^2(u) \mathcal U_{kl}^2(u).
\end{equation} 

For non-positive curvature, this expression as well as the volume $V$ need regularization. We will do so by restricting the $u$-integral to the interval $(0,R)$ and then we will take the limit $R\to\infty$; also the volume of a sphere of radius $R$ is $V_R=4\pi\int_0^R\diff u r^2(u)$. 
In this case we obtain that the total particle density is
\begin{align}
    n&=\frac{1}{V_R}\int_0^R\diff u r^2(u) \int \diff\mu_k|\beta_k|^2\sum_{l=0}^\infty (2l+1)\mathcal U_{kl}^2(u)
    \nonumber\\
    &=\frac{2}{\pi V_R}\int_0^R\diff u r^2(u)  \int \diff\mu_k|\beta_k|^2=\frac{1}{2\pi^2}   \int \diff\mu_k|\beta_k|^2,
\end{align}
where in the first equality we have performed the sum over $q$ and  in the second we have used the summation rule \eqref{eq:sumruleU}.

For positive curvatures, there is no need for regularization. In this case
\begin{align}
    n&=\frac{1}{V}\int_0^{\pi/\sqrt{\abs{\kappa}}}\diff u r(u)^2  \int \diff\mu_k|\beta_k|^2\sum_{l=0}^{\bar k}(2l+1)\mathcal U_{kl}^2(u)
    \nonumber\\
    &=\frac{2}{\pi V}\int_0^{\pi/\sqrt{\abs{\kappa}}}\diff u r(u)^2  \int \diff\mu_k|\beta_k|^2=\frac{1}{2\pi^2}  \int \diff\mu_k |\beta_k|^2,
\end{align}
where again in the first equality we have performed the sum over $q$,  in the second we have used the addition rule \eqref{eq:sumruleU}, and in the last we have used the expression for the volume $V=4\pi\int_0^{\pi/\sqrt{\abs{\kappa}}}\diff u r^2(u)$.

The ambiguity in the election of a quantum theory starting from a given classical theory is natural to all quantum field theories, even in Minkowski spacetime. In the latter scenario, however, one typically asks the quantum theory to respect the Poincaré symmetry of the classical theory. This selects a unique basis of solutions, leading to the Minkowski vacuum. However, when time-translational invariance is broken, as it may be the case when the geometry changes with time, the group of symmetries of the classical theory is not enough to restrict the possible choices of quantization. Therefore, the notions of vacuum and particle become ambiguous, and quantum vacuum ambiguities are inherent to quantum field theories in curved spacetime \cite{Alvarez2023b}.

\subsection{Cosmological particle production as an \textit{in-out} process}
\label{subsec:inout}

To understand cosmological particle production, let us analyze the following example. Consider the scenario in which the spacetime undergoes certain expansion between $\eta_{\text{i}}$ and $\eta_{\text{f}}$, but is static for $\eta\leq\eta_{\text{i}}$ and \mbox{$\eta\geq\eta_{\text{f}}$}, such that the frequency of the mode equation~\eqref{eq:Frequency} acquires the constant value~$\omega_{k, \text{i}}$ and $\omega_{k, \text{f}}$, respectively, and is time-dependent in the period $\eta_{\text{i}} < \eta < \eta_{\text{f}}$. Observers living in the \textit{in} ($\eta \leq \eta_{\text{i}}$) and \textit{out} ($\eta \geq \eta_{\text{f}}$) regions will have a preferred notion of vacuum, corresponding to selecting plane waves as modes for the field expansion \eqref{eq:FieldExpansion}, similarly as canonical quantization proceeds in Minkowski spacetime \cite{Birrell1982}. Thus, the natural notion of vacuum for an observer living in the \textit{in} region will correspond to modes that behave as
\begin{equation}
    v_k = \frac{1}{\sqrt{2\omega_{k, \text{i}}} }e^{-i\omega_{k, \text{i}} \eta}, \quad \text{for} \quad \eta \leq \eta_{\text{i}},
\label{eq:InModeIC}
\end{equation}
and similarly for an observer living in the \textit{out} region,
\begin{equation}
    u_k = \frac{1}{\sqrt{2\omega_{k, \text{f}}} }e^{-i\omega_{k, \text{f}} \eta},\quad \text{for} \quad \eta \geq \eta_{\text{f}},
\label{eq:OutModeIC}
\end{equation}
where the modes are normalized according to the Wronskian \eqref{eq:WronskianModes}. Expanding the field in terms of $v_k$ or $u_k$ and its complex conjugates leads to two particular sets of creation and annihilation operators defining the \textit{in} and \textit{out} vacua, respectively. In particular, for~\mbox{$\eta \geq \eta_{\text{f}}$},~$v_k$ will not correspond to the natural notion of vacuum for an observer living in that region, which is in fact associated with the mode $u_k$, but will behave instead as some linear combination of positive and negative frequency plane waves, as \cref{eq:ModeEquation} dictates in that regime in this example. Since each mode, together with its conjugate, is a basis of solutions, the modes~$v_k$ and $u_k$ are related by a Bogoliubov transformation of the form \eqref{eq:BogoliubovTransformationModes}, which in general will involve a non-vanishing $\beta_k$ coefficient. This indicates that the notion of vacuum before and after the expansion is not the same, or, put differently, that the expansion of the geometry produces particles---measured by an observer after this process---out of the vacuum, as understood by an observer before the change of spacetime. 

However, spacetime is always expanding. Nevertheless, at some times the expansion can be very adiabatic. The effect that non-adiabaticity of spacetime has in particle production can be estimated by examining the dimensionless function $\mathcal{C}_k$ \cite{Birrell1982,Alvarez2025,Duran2026},
\begin{equation} 
    \mathcal{C}_k(\eta)\equiv\abs{\omega_k^{\prime}(\eta)/\omega_k^2(\eta)}.
    \label{eq:AdiabaticCoefficient}
\end{equation}
In a scenario in which spacetime expansion is very slow ($\abs{\mathcal{C}_k(\eta)} \ll 1$) at early and late times---which can be thought of as approximately \textit{in} and \textit{out} regions---, a suitable vacuum choice is the so-called adiabatic vacuum, characterized by the modes with initial conditions\footnote{This is related to approximate solutions to the mode equation in the WKB formalism \cite{Bender1999}.}
\begin{equation}
    v_k(\eta_{\text{ad}}) = \frac{1}{\sqrt{2\omega_k(\eta_{\text{ad}})}}, \quad 
    v_k^{\prime}(\eta_{\text{ad}}) = -\frac{1}{\sqrt{2\omega_k(\eta_{\text{ad}})}}\left(i\omega_k(\eta_{\text{ad}}) + \frac{1}{2}\frac{\omega_k^{\prime}(\eta_{\text{ad}})}{\omega_k(\eta_{\text{ad}})}\right).
\label{eq:AdiabaticVacuum}
\end{equation}
This is an appropriate vacuum prescription \textit{at each time $\eta_{\text{ad}}$} as long as the adiabaticity condition is satisfied.

In order to make the right choice of vacuum for the physical situation in hand, we introduce the background geometry in the next section. 

\section{De Sitter spacetime as a toy model of inflation}
\label{sec:deSitter}

The early dynamics of a FLRW universe with spatial curvature $\kappa$ and scale factor~$a$ are described by the Friedmann equation 
\begin{equation}
    H^2 = \frac{8\pi}{3\mpl^2}\rho - \frac{\kappa}{a^2},
\label{eq:FriedmannEquation}
\end{equation}
with the energetic content of the universe dominated by the inflaton field $\phi$, and $H = \dot{a}/a$ is the Hubble rate. That is, $\rho=\dot{\phi}^2/2+V(\phi)$, where $V(\phi)$ is the inflaton potential. Therefore, the Hubble rate acquires the form
\begin{equation}
    H^2 = \frac{4\pi}{3\mpl^2}\left[\dot{\phi}^2 + 2V(\phi)\right] - \frac{\kappa}{a^2}.
\label{eq:InflatonFriedmann}
\end{equation}
Assuming that the derivative of the inflaton is negligible compared to the potential---i.e., within the slow-roll approximation---the above expression can be approximated as
\begin{equation}
    H^2(t) \simeq \frac{8\pi}{3\mpl^2}V(\phi) - \frac{\kappa}{a^2(t)}.
\end{equation}
The inflaton potential approaches a constant value $V_0$ in the past, and
\begin{equation}
    H^2(t) \to \frac{8\pi}{3\mpl^2}V_0 - \frac{\kappa}{a^2(t)} = H_0^2 - \frac{\kappa}{a^2(t)}, 
\label{eq:deSitterHubbleRate}
\end{equation}
with $H_0$ denoting the Hubble rate in this limit for $\kappa=0$. This is precisely the Friedmann equation for a de Sitter universe with spatial curvature $\kappa$, whose energetic content is characterized by a cosmological constant~\mbox{($\rho \simeq \text{constant}$)}. Therefore, de Sitter spacetime is a good approximation to (the early regime of) an inflationary model characterized by a very slow-roll of the inflaton\footnote{Although of course, it does not capture the oscillatory dynamics in reheating \cite{Cembranos2023}.}. In order to perform analytic calculations throughout, we will study the influence of spatial curvature in the production of particles assuming a de Sitter geometry for the whole inflationary stage.

\subsection{Exit of inflation, reheating and radiation dominance}
\label{subsec:exit}

When the slow-roll regime ends---and therefore the de Sitter description is no longer valid---, the inflaton field~$\phi$ decays and starts oscillating around the minimum of the potential~$V(\phi)$. This is the typical scenario for chaotic, single field inflation governed by, e.g., a Starobinsky or quadratic potential. After that, reheating begins and the energy density of the inflaton is transferred to other particles via their interactions. At this point, the scale factor ceases to behave as a (quasi-)exponential, $a(t) \sim e^{H_0 t}$, and acquires the typical form of the scale factor of a matter-dominated universe, $a(t) \sim t^{2/3}$: Inflation ends and reheating starts. Eventually, the expansion of the geometry slows down even more, and the radiation-dominated epoch begins. Interestingly, the rate of production of particles becomes increasingly negligible from the late reheating period onwards. We model this transition from inflation to reheating and radiation dominance in two different ways.

\begin{figure}[t]
    \centering
    \includegraphics[width=0.45\textwidth]{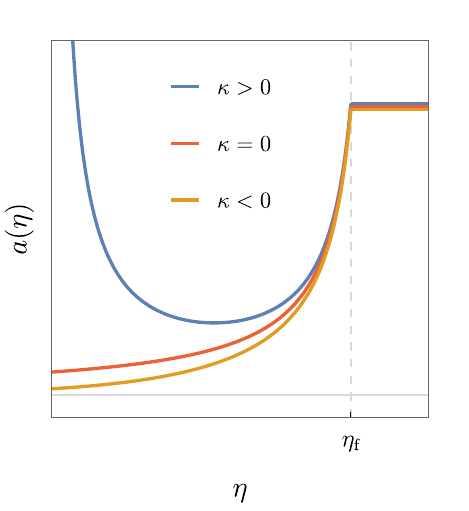}
    \caption[De Sitter scale factors for the three cosmological patches]{Time dependence of the scale factor for the three different spatial curvatures, assuming an instantaneous transition to a static geometry at $\eta_{\text{f}}$. The scale factor is normalized so that the value it takes at the end of inflation is \mbox{$a(\eta_{\text{f}}) = 1$}.}
    \label{fig:ScaleFactors}
\end{figure}

As a first approximation, we consider that inflation, i.e., the de Sitter-like phase, ends at a given conformal time $\etaf$, where the scale factor becomes static, and reheating happens instantaneously at that point. In \cref{fig:ScaleFactors}, we show the time dependence of the scale factor as function of conformal time. Since spacetime is static after $\etaf$, an observer living in this region will have a preferred notion of vacuum, as discussed above. The frequency of the mode equation \eqref{eq:Frequency} becomes constant, and the basis defining the natural vacuum for an observer living at $\eta \geq \etaf$ are the \textit{out} modes $u_k(\eta)$ of the form \eqref{eq:OutModeIC} with positive frequency~\mbox{$\omega_{\text{f}}=\omega(\eta_{\text{f}})$}, as discussed before. We will refer to this vacuum notion as the instantaneous vacuum.

Alternatively, instead of assuming an instantaneous transition to a static geometry, one could also consider a smooth transition. In this case, one would need to take into account the specific evolution of the geometry and make an appropriate vacuum choice. However, in the line of the analysis in Ref. \cite{Alvarez2023}, it can be seen that a sufficiently smooth transition beginning at some time $\eta$ leads to the same number of particles as the one obtained by using an adiabatic vacuum prescription with initial conditions at that same time $\eta$. In this sense, one can find an order of adiabatic vacuum compatible with a transition of certain smoothness. We will not explicit any specific transition, but compute instead particle production using the adiabatic prescription \eqref{eq:AdiabaticVacuum}, which encodes the evolution of the geometry.

Once the notion of \textit{out} vacuum after inflation is selected, it remains to choose the \textit{in} modes. For this, we need to study the quantization of the scalar field in the de Sitter epoch, to which we devote the next section.

\section{Quantization in de Sitter spacetime}
\label{sec:desitter}

We have seen that at the onset of inflation, the geometry approaches that of de Sitter. Moreover, considering an exact de Sitter geometry during the full inflationary phase is a well justified assumption and serves as a first, simple model. Regardless of whether one makes this approximation or not, in order to choose a suitable notion of \textit{in} vacuum, one needs to answer the question of how to quantize fields in de Sitter spacetime. Crucially, one can define a preferred notion of vacuum in this geometry (for the massive scalar case), the so-called Bunch-Davies vacuum~\mbox{\cite{Birrell1982,Allen1985,Mottola1986}}.

In the following, we will review how the three different cosmological cases under study (positive, negative, and vanishing $\kappa$) are realized by different choices of coordinates, covering different regions, in a de Sitter geometry.

\subsection{Spatially curved inflationary cosmologies as de Sitter patches}

\begin{figure}
    \centering
    \scalebox{0.95}{
        \begin{overpic}[scale=0.45]{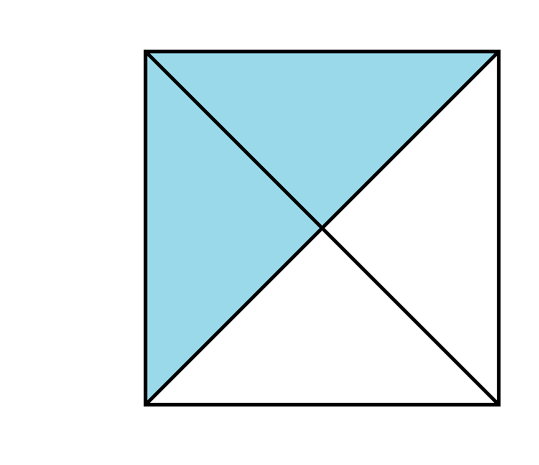}
        \put(17, 80){(a)}
        \put(17, 33){\rotatebox{90}{$r=0$}}
        \put(50, 80){$t=\infty$}
        \put(30, 22){\rotatebox{45}{$t=-\infty$}}
        \put(60, 52){\rotatebox{45}{$r=\infty$}}
        \end{overpic}
        \hspace*{-4mm}
        \begin{overpic}[scale=0.45]{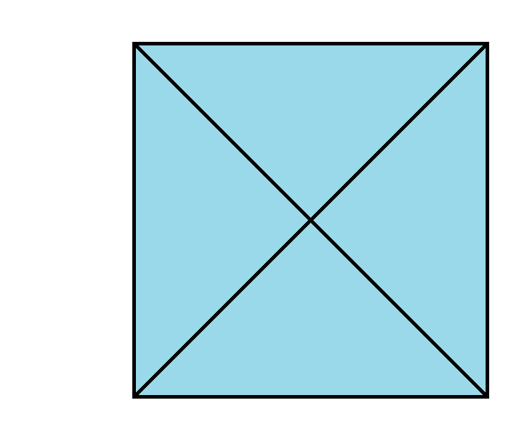}
        \put(17, 81){(b)}
        \put(17, 34){\rotatebox{90}{$r=0$}}
        \put(46, 81){$t=\infty$}
        \put(46, 2){$t=-\infty$}
        \put(96, 33){\rotatebox{90}{$r=\pi$}} 
        \end{overpic}
        \begin{overpic}[scale=0.45]{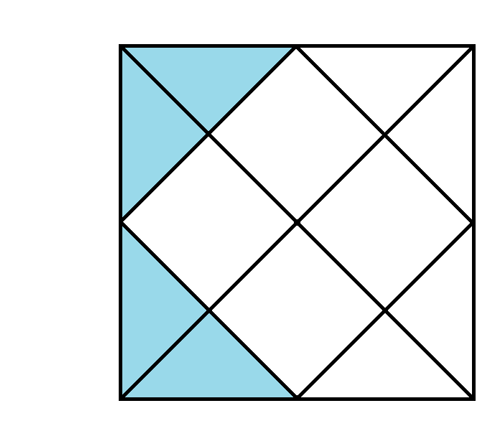}
        \put(15, 84){(c)}
        \put(14, 34){\rotatebox{90}{$r=0$}}
        \put(47, 83){$t=\infty$}
        \put(47, 2){$t=-\infty$}
        \put(33, 48){\rotatebox{45}{$t=0$}}
        \put(33, 38){\rotatebox{-45}{$r=\infty$}}
        \end{overpic}            
        }
\caption[De Sitter Penrose diagrams]{Penrose diagrams for the different coordinate charts in de Sitter: (a) flat, (b) closed, and (c) open coordinates, in terms of cosmological time $t$. The shadowed region corresponds to the covering of the corresponding coordinates, which are denoted with the same symbol in each case.}
\label{fig:PenroseDiagrams}
\end{figure}
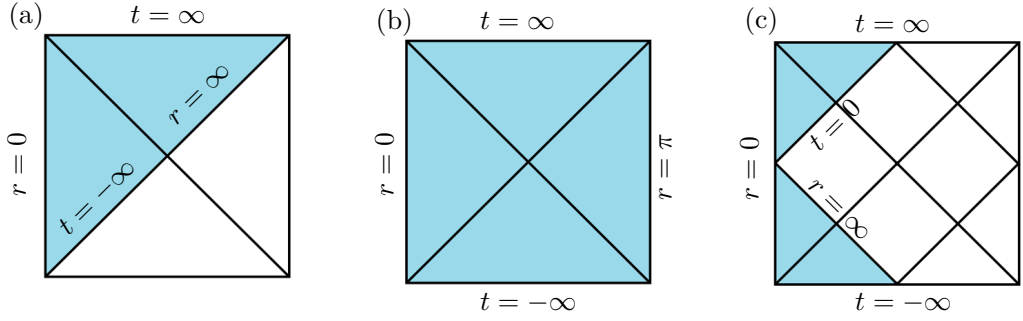

From \cref{eq:deSitterHubbleRate} we can obtain the scale factor $a(\eta)$ as a function of conformal time,
\begin{equation}\displaystyle
    a(\eta) = \begin{cases}
    \frac{\sqrt{\abs{\kappa}}}{H_0}\sin^{-1}{\left(\sqrt{\abs{\kappa}}\eta\right)} \quad &\text{for} \quad \kappa > 0,\\
    -\frac{1}{H_0\eta} \quad &\text{for} \quad \kappa = 0,\\
    \frac{\sqrt{\abs{\kappa}}}{H_0}\sinh^{-1}{\left(\sqrt{\abs{\kappa}}\eta\right)} \quad &\text{for} \quad \kappa < 0,
    \end{cases}
\label{eq:ScaleFactors}
\end{equation}
where $\eta \in (-\pi/\sqrt{\abs{\kappa}}, 0)$ in the first case and $\eta \in \left(-\infty, 0\right)$ in the rest. On the other hand, the Ricci scalar in this geometry is constant and given by~\mbox{$R = 12H_0^2$}, which obviously does not depend on the patch considered. The corresponding line element can be written for the different values of~$\kappa$ as in \cref{eq:GeneralLineElement}, where the scale factor has to be taken from~\cref{eq:ScaleFactors}. This is precisely the form that the line element takes in a de Sitter geometry when using closed, flat, or open coordinate charts (corresponding to the cases~\mbox{$\kappa>0$}, \mbox{$\kappa=0$} and $\kappa<0$, respectively)~\mbox{\cite{Birrell1982,Mottola1986}}. The Penrose diagrams for the different coordinate charts are displayed in Fig. \ref{fig:PenroseDiagrams}.

With this, we can explicitly write the mode equations for the three different coordinate choices in de Sitter, or, equivalently, for the three different FLRW cosmologies (closed, flat or open). They are given by
\begin{equation}
    v_k^{\prime\prime}(\eta) + \omega_k^2(\eta)v_k(\eta)=0,
\label{eq:ModeEquations}
\end{equation}
with  
\begin{equation}
    \omega_k^2(\eta)=\begin{cases}
    \left(k+\sqrt{\kappa}\right)^2 + \abs{\kappa}\mu^2/\sin^2{\left(\sqrt{\abs{\kappa}}\eta\right)}, \quad &\text{for} \quad \kappa>0, \\
    k^2 + \mu^2/\eta^2, \quad &\text{for} \quad \kappa=0, \\
    k^2 + \abs{\kappa}\mu^2/\sinh^2{\left(\sqrt{\abs{\kappa}}\eta\right)}, \quad &\text{for} \quad \kappa<0,
    \end{cases}
\label{eq:Frequencies}
\end{equation}
where $\mu$ is a dimensionless, effective mass defined as
\begin{equation}
    \mu^2 = m^2/H_0^2 + 12\left(\xi - 1/6\right).
\end{equation}

We have already mentioned that de Sitter spacetime has a preferred notion of vacuum, the Bunch-Davies vacuum, which is associated with creation and annihilation operators that were the coefficients of the expansion of the fields in a particular basis of the solutions of \cref{eq:ModeEquations} before quantization. In the following subsection we discuss which particular solutions of the mode equations lead to an expansion in terms of the operators defining said vacuum, for each of the coordinate charts considered, associated with different cosmological scenarios.

\subsection{Bunch-Davies vacuum}

Let us analyze the choice of Bunch-Davies modes for the different de Sitter patches. 

\subsubsection*{Flat patch}
\label{subsubsec:flatBD}

We study first the flat coordinate chart, which is typically considered in the context of cosmology, since it is often assumed that spatial sections are flat. The general solution to the mode equation \eqref{eq:ModeEquations} can be written in this case as
\begin{equation}
    v_k(\eta) = A_k H_{\nu}^{(1)}(k\abs{\eta})+B_k H_{\nu}^{(2)}(k\abs{\eta}),
\end{equation}
with $\nu = \sqrt{1/4-\mu^2}$. Note that in the limit $k\abs{\eta} \gg 1$, the frequency \eqref{eq:Frequencies} becomes constant, and therefore one expects that the modes $v_k$ behave as $e^{-i\omega_k\eta}$ in this limit. Making use of the asymptotic behavior of Hankel functions, one finds that the coefficients of the linear combination that becomes a positive frequency solutions for~\mbox{$k \abs{\eta} \gg 1$} are those that fulfill that \cite{Birrell1982,Mukhanov2007}
\begin{equation}
    A_k \overset{k\abs{\eta}\gg 1}{\longrightarrow} \frac{1}{2}\sqrt{\frac{\pi}{k}} e^{i\frac{\pi}{2}\left(\nu + \frac{1}{2}\right)}, \qquad B_k \overset{k\abs{\eta}\gg 1}{\longrightarrow} 0.
\end{equation}
If we extend these values to the constants for all wavenumbers~$k$, the resulting mode takes the form
\begin{equation}
    v_k(\eta) = \frac{\sqrt{\pi \abs{\eta}}}{2} e^{i\frac{\pi}{2}\nu}H_{\nu}^{(1)}(k\abs{\eta}),
\label{eq:flatBD}
\end{equation}
where we neglected an overall constant phase. The modes \eqref{eq:flatBD} define the Bunch-Davies vacuum, written in flat coordinates.

\subsubsection*{Closed patch}

In this case, the general solution to the mode equation \eqref{eq:ModeEquations} is given by
\begin{equation}
    v_k(\eta) = \abs{\kappa}^{-1/4}\sin^{1/2}{\left(\sqrt{\abs{\kappa}}\eta\right)}\Bigg\{A_k P^{\nu}_{{\bar k}+1/2}\left[-\cos{\left(\sqrt{\abs{\kappa}}\eta\right)}\right]+B_k Q^{\nu}_{{\bar k}+1/2}\left[-\cos{\left(\sqrt{\abs{\kappa}}\eta\right)}\right]\Bigg\},
\end{equation}
where $P_k^{\nu}$ and $Q_k^{\nu}$ are the associated Legendre polynomials, and $\nu$ is defined as in the flat case. Similar to what we did before, we examine the behavior of the modes in the limit $k \to \infty$, for which the frequency becomes time-independent. This time, we take for the modes that behave as positive frequency plane waves the values of the constants \cite{Birrell1982,Allen1985,Mottola1986}
\begin{equation}
    A_k = \frac{\sqrt{\pi}}{2}\left({\bar k}+1\right)^{-\nu}e^{i\left[\left({\bar k}+1\right)\pi - \frac{\pi}{4} + \frac{\nu\pi}{2}\right]}, \qquad B_k = \frac{2i}{\pi} A_k.
\end{equation}
Finally, the mode reads
\begin{equation}
    v_k(\eta) = \abs{\kappa}^{-1/4}\sin^{1/2}{\left(\sqrt{\abs{\kappa}}\eta\right)}A_k\Bigg\{P^{\nu}_{{\bar k}+1/2}\left[-\cos{\left(\sqrt{\abs{\kappa}}\eta\right)}\right]+\frac{2i}{\pi}Q^{\nu}_{{\bar k}+1/2}\left[-\cos{\left(\sqrt{\abs{\kappa}}\eta\right)}\right]\Bigg\},
\label{eq:closedBD}
\end{equation}
It can be checked that the two-point function calculated in the stated defined by Eqs.~\eqref{eq:closedBD} and~\eqref{eq:flatBD} yields the same result \cite{Chernikov1968,Tagirov1973, Schomblond1976,Bunch1978}. Therefore, both modes define the same vacuum, i.e., Bunch-Davies, this time expressed in closed cosmological coordinates.

\subsubsection*{Open patch}

Now, the general solution to the \cref{eq:ModeEquations} is given by
\begin{equation}
    v_k(\eta) = \abs{\kappa}^{-1/4}A_k P^{i{\bar k}}_{\nu}\left[\coth{\left(-\sqrt{\abs{\kappa}}\eta\right)}\right] + B_k P^{-i{\bar k}}_{\nu}\left[\coth{\left(-\sqrt{\abs{\kappa}}\eta\right)}\right].
\end{equation}
The analysis is more involved in this case. The open coordinates can be used to cover the left or the right open patches (only the left one is highlighted in figure~\ref{fig:PenroseDiagrams}). One can define modes in either patch and extend them analytically to the opposite region. Then, the two following linear combinations of those extended modes (left to the right and right to the left) expand the field in operators annihilating the Bunch-Davies vacuum~\mbox{\cite{Sasaki1994,Maldacena2013,Kanno2014}},
\begin{equation}
    v_{k}^{(\pm)} = \frac{v_k^{(\text{R})}\pm v_k^{(\text{L})}}{\left(1\pm e^{-\pi {\bar k}+i\left(\nu-1/2\right)\pi}\right)\Gamma(\nu+1/2+i{\bar k})},   
\end{equation}
where $v_k^{(R)}$ and $v_k^{(L)}$ are given by
\begin{equation}
\begin{split}
    v_k^{(R)} &= \frac{i}{\pi}\left(1+e^{-\pi {\bar k}+i\left(\nu-1/2\right)\pi}\right)\left(1-e^{\pi {\bar k} - i\left(\nu-1/2\right)\pi}\right)Q^{i{\bar k}}_{\nu-1/2}\left[\coth{\left(-\sqrt{\abs{\kappa}}\eta\right)}\right] ,\\
    v_k^{(L)} &= \left(1+e^{-\pi {\bar k}+i\left(\nu-1/2\right)\pi}\right) P^{i{\bar k}}_{\nu-1/2}\left[\coth{\left(-\sqrt{\abs{\kappa}}\eta\right)}\right],
\end{split}
\end{equation}
up to a normalization factor. The linear combinations can be rewritten as
\begin{equation}
\begin{split}
    v_{k}^{(\pm)} = \frac{\pm N_k}{2\sinh{\left(\pi {\bar k}\right)}}\Bigg\{&\frac{e^{\pi {\bar k}} \mp e^{-i\pi\left(\nu-1/2\right)}}{\Gamma(\nu+1/2+i{\bar k})}P^{i{\bar k}}_{\nu-1/2}\left[\coth{\left(-\sqrt{\abs{\kappa}}\eta\right)}\right] \\ 
    &-\frac{e^{-\pi {\bar k}} \mp e^{-i\pi\left(\nu-1/2\right)}}{\Gamma(\nu+1/2-i{\bar k})}P^{-i{\bar k}}_{\nu-1/2}\left[\coth{\left(-\sqrt{\abs{\kappa}}\eta\right)}\right]\Bigg\},
\label{eq:openBD}
\end{split}
\end{equation}
where $N_k$ is given by
\begin{equation}
\begin{split}
    N_k^2=2\pi\abs{\kappa}^{-1/2}&\sinh{(\pi {\bar k})}\Bigg\{\abs{\Gamma(\nu+1/2+i{\bar k})}^{-2}\left(e^{2\pi {\bar k}}+1 \mp e^{\pi {\bar k}}2\cos{[\pi\left(\nu-1/2\right)]}\right)\\
    &\hspace{0.5cm}-\abs{\Gamma(\nu+1/2-i{\bar k})}^{-2}\!\left(e^{-2\pi {\bar k}}\!+\!1 \mp e^{-\pi {\bar k}}2\cos{[\pi\left(\nu-1/2\right)]}\right)\!\Bigg\}^{-1}.
\end{split}
\end{equation}
The field expansion \eqref{eq:FieldExpansion} is written as
\begin{equation}
    \chi(\eta, \vec{u}) = \int \dd \mu_{klq}\sum_{s=\pm} \left[a_{klq}^{(s)}\mathcal{H}_{klq}(\vec{u})v_k^{(s)}(\eta) + a_{klq}^{*, \, (s)}\mathcal{H}_{klq}^*(\vec{u})v_k^{*,(s)}(\eta)\right].
\end{equation}
The three particular solutions \eqref{eq:flatBD}, \eqref{eq:closedBD} and \eqref{eq:openBD} are associated with an expansion of the field operator corresponding to creation and annihilation operators defining the Bunch-Davies vacuum \cite{Birrell1982,Mottola1986,Sasaki1994}. We will assume that this is the initial state of our system at the onset of inflation. In other words, the initial condition for our mode equation will be given by~\cref{eq:flatBD,eq:closedBD}, and \eqref{eq:openBD} depending on the case we are studying. 

\section{Particle production}
\label{sec:particleproduction}

It only remains to compare the Bunch-Davies modes with those associated with the vacuum notion after the exit of inflation through~\cref{eq:BogoliubovCoefficientsWronskian}, be it plane waves of the form~\eqref{eq:OutModeIC} or zeroth-order adiabatic modes with initial conditions~\eqref{eq:AdiabaticVacuum}, which we evaluate at~$\eta_{\text{f}}$. In the following, we show results for the number density of produced particles in the different cases, as well as the corresponding relic abundance today in the case that the spectator field $\varphi$ constitutes dark matter.

If these gravitationally produced scalar particles are considered to be dark matter, one must ensure that their abundance is compatible with observations. The physical density of produced particles is related to the comoving density $n=\int \dd \mu_{klq} \abs{\beta_k}^2/(2\pi^2)$ only by the scale factor. Assuming that the scalar field is non-interacting, which a typical scenario in which gravitational production becomes relevant, as it cannot reach thermal equilibrium, the evolution of the density of created particles from $\etaf$ until today will be dictated solely by the dilution due to the isentropic expansion of the background. The predicted abundance can be written in terms of the background radiation temperature, assuming a standard cosmological evolution, as
\begin{equation}
    \Omega (m, \xi) = \frac{8\pi}{3\mpl^2H^2_{\text{today}}}\frac{g_{*S}^{\text{today}}}{g_{*S}^{\text{rh}}}\left(\frac{T_{\text{today}}}{T_{\text{rh}}}\right)^3 m \, \frac{n(m, \xi)}{a_{\text{rh}}^3},
\label{eq:Abundance}
\end{equation}
where $T_{\text{today}}$ and $T_{\text{rh}}$ are the radiation temperature today and at the end of reheating, respectively, and $g^{\text{today}}_{*S}$ and $g^{\text{rh}}_{*S}$ are the corresponding relativistic degrees of freedom. In the above expression, we have assumed that these particles are non-relativistic at present, so that their energy density is dominated by their rest mass, making the abundance proportional to the product of the mass and the number density.

\begin{figure}[t!]
    \centering
    \includegraphics[width=0.5\textwidth]{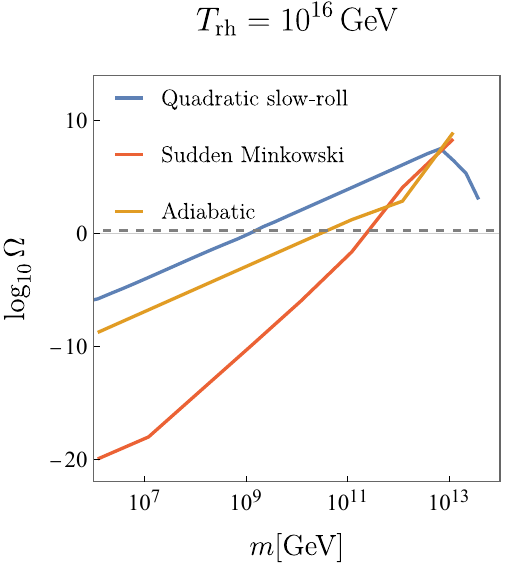}
    \caption{Abundance for today-non-relativistic particles by assuming flat quadratic inflation (blue), de Sitter with instantaneous vacuum (orange), and de Sitter with adiabatic vacuum (yellow). The mass of the spectator field is given in GeV. The dashed line denotes the observed dark matter abundance.} 
    \label{fig:TotalDensityFlat}
\end{figure}

Let us first compare production in the flat ($\kappa=0$) case, for both choices of vacuum after inflation---instantaneous and adiabatic---with that of a single-field inflationary model with quadratic potential from Refs. \cite{Cembranos2023,Cembranos2025a}. We performed the computations for a Hubble rate of $H_0 \simeq 7.4 \times 10^{13} \, \text{GeV}$ in the flat scenario. Figure \ref{fig:TotalDensityFlat} shows the abundance of produced particles as function of the spectator field mass $m$, diluted until today, for a reheating temperature of~\mbox{$T_{\text{th}}=10^{16} \, \text{GeV}$}. The horizontal dashed line denotes the observed dark matter abundance.

We can see that production of particles is of the same order in all cases for values of the mass of the field around the inflaton mass $m_{\phi}=1.2 \times 10^{13} \, \text{GeV}$ for the inflationary model in Ref. \cite{Cembranos2023}. However, we find large deviations for smaller masses. Interestingly, the adiabatic vacuum prediction evolves with the mass similarly to the quadratic inflation case, whereas it coincides with the instantaneous prediction in the inflaton mass regime. The latter is due to the fact that the difference between adiabatic and instantaneous modes vanishes for large mass. Indeed, the derivative of the adiabatic mode in \cref{eq:AdiabaticVacuum} contains the term
\begin{equation}
    \frac{\omega_{k,\text{dS}}^{\prime}}{\omega_{k, \text{dS}}^2} = \frac{aa^{\prime}m^2}{\left(k^2+a^2m^2\right)^{3/2}} = -\frac{m^2}{H_0^2\left(k^2\eta^2 + m^2/H_0^2\right)^{3/2}}.
\label{eq:AdiabaticCoefficientdS}
\end{equation}
For UV modes, $k\gg am = m/(H_0\eta)$, the fraction \eqref{eq:AdiabaticCoefficientdS} grows with $m$, but it is nevertheless very small.  For wavenumbers for which production is important, those fulfilling $k\lesssim m/(H_0\eta)$, the adiabatic coefficient is of order one. Note that the opposite limit can be reached for sufficiently high masses,~\mbox{$k \ll m/(H_0\eta)$}, and in this situation $\omega_{k,\text{dS}}^{\prime}/\omega_{k, \text{dS}}^2 \sim H_0/m$. This is the reason why the adiabatic and instantaneous predictions coincide for large masses, and deviate more and more for smaller masses.

\begin{figure}[t!]
    \centering
    \includegraphics[width=0.5\textwidth]{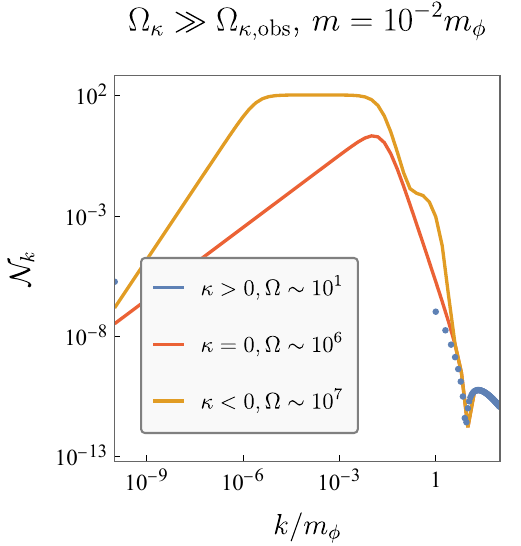}
    \caption{Spectra of produced particles with different spatial curvature in de Sitter inflation using the adiabatic vacuum prescription, for a large value of the curvature abundance $\Omega_\kappa$ (corresponding to $|\kappa| = m_\phi$) and $m = 10^{-2}m_\phi$. Here, $\mathcal N_k = (k/m_\phi)^2|\beta_k|^2$ for $\kappa \leq 0$, whereas $\mathcal N_k = (k + \sqrt{|\kappa|})^2/m_{\phi}^2|\beta_k|^2$ for $\kappa > 0$ (see also Eq. (2.13)). The abundances $\Omega$ in the legend denote the associated relic abundance obtained by using Eq. (5.1) with $T_{\text{rh}} = 10^{15}$ GeV. Note that the spectrum for positively curved scenarios is discrete; in particular, the blue point at the far left of the spectrum corresponds to the $k=0$ mode. Positive and negative curvature lead to a smaller and larger abundance than in the flat case, respectively. The differences can be of several orders of magnitude.}
    \label{fig:SpectraCurvedLargeK}
\end{figure}

\begin{figure}[t!]
    \centering
    \includegraphics[width=0.5\textwidth]{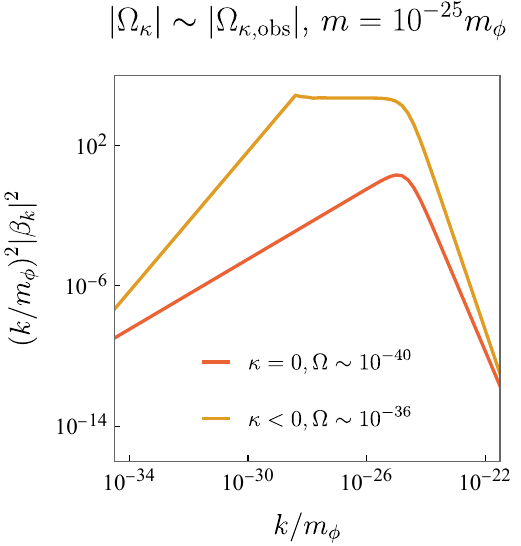}
    \caption{Spectra of produced particles with zero and negative spatial curvature in de Sitter inflation using the adiabatic vacuum prescription for $\abs{\kappa} = 10^{-40}m_{\phi}$ and~\mbox{$m=10^{-25}m_{\phi}$}. It shows that significant deviations in the spectra are obtained for fields in the ultra-light regime even considering curvature values compatible with current observations. Although the total abundance for such light fields represents a subleading contribution to the dark matter density, these results reveal a strong sensitivity of the production mechanism to the global geometry of the universe. The abundance is obtained using~\mbox{$T_{\text{rh}}=10^{15} \, \text{GeV}$}. Note that for such light fields and small spatial curvature, the spectrum of produced particles in the case of positive curvature is negligible compared to the other two cases shown in the Figure.} 
    \label{fig:SpectraCurvedSmallK}
\end{figure}

One can also extract another lesson from the analysis of \cref{eq:AdiabaticCoefficientdS}. At the beginning of inflation, when the scale factor is very small, the condition~\mbox{$k\lesssim m/(H_0\eta)$} for production is only fulfilled by arbitrarily small wavenumbers (or large wavelengths). Consequently, production of higher $k$ happens after a sufficiently long expansion period. On the other hand, it seems that the mass dependence of the abundance is not sensitive to the mode evolution during inflation for $m\lesssim H_0 \sim m_{\phi}$, as can be observed when comparing the quadratic and adiabatic curves in~\cref{fig:TotalDensityFlat}. Even more, the fact that the instantaneous prediction disagrees with the other two leads to the idea that this behavior depends mainly on the choice of vacuum.

Let us now examine the spectra of produced particles for the different values of spatial curvature $\kappa$. In Fig. \ref{fig:SpectraCurvedLargeK}, we show the resulting spectra in the context of de Sitter inflation when choosing adiabatic conditions for the \textit{out} vacuum. We can see that production is enhanced in an open universe with respect to the flat case, whereas it yields a smaller number density in a closed universe. This is due to the fact that the curvature of the spatial sections affects the evolution of the modes, and therefore the number of particles produced. The effect is more pronounced for small masses, where the curvature of the spatial sections is more relevant and the~\mbox{$\kappa$-dependent} terms in the mode equations \eqref{eq:ModeEquations} become more important. 

From observational constraints, we know that the abundance of curvature in the nucleosynthesis era had to be smaller than \mbox{$\Omega_{\kappa, \text{nuc}} = 10^{-16}$}. Assuming a radiation-dominated evolution backwards in time until the end of reheating, one can deduce
\begin{equation}
    \Omega_{\kappa, \text{rh}} = \Omega_{\kappa, \text{nuc}}\left(\frac{T_{\text{nuc}}}{T_{\text{rh}}}\right)^2 \sim 10^{-16}\left(\frac{T_{\text{nuc}}}{T_{\text{rh}}}\right)^2,
\label{eq:ObservedCurvatureAbundance}
\end{equation}
where $T_{\text{nuc}} \simeq 10^{-3} \, \text{GeV}$ and $T_{\text{rh}}$ are the nucleosynthesis and reheating temperatures, respectively. The latter is treated as a free parameter, and considered to lie somewhere between~$T_{\text{nuc}}$ and the Planck scale at $10^{19} \, \text{GeV}$. On the other hand, the abundance of curvature at the end of reheating can be obtained from 
\begin{equation}
    \Omega_{\kappa, \text{rh}} = -\frac{\kappa}{a_{\text{rh}}(T_{\text{rh}})H(T_{\text{rh}})},
\label{eq:CalculatedCurvatureAbundance}
\end{equation}
where both the scale factor and the Hubble rate can be written in terms of the reheating temperature, assuming matter domination until the end of reheating---as is the case for inflationary potentials that behave quadratically around its minimum. From \cref{eq:ObservedCurvatureAbundance,eq:CalculatedCurvatureAbundance} one can obtain the value of $\kappa$ for which the curvature abundance is compatible with observations at each value of $T_{\text{rh}}$ considered. It turns out that compatible values of $\kappa$ are far from the one used in \cref{fig:SpectraCurvedLargeK}. One must therefore go to very low wavenumbers (large enough wavelengths) so that the small spatial curvature is noticed by the corresponding modes. On the other hand, the spectra shifts towards lower values of $k$ with decreasing mass $m$. Therefore, if one aims at observing differences in production due to the curvature, one must consider very low masses, such that production of particles is important at wavenumbers of the order of $\abs{\kappa}$. This is precisely what is shown in \cref{fig:SpectraCurvedSmallK}, where there are differences of four orders of magnitude in the abundance between the different curvatures. However, for curvature values within current observational bounds, the associated abundance remains small. We therefore find that while spatial curvature fundamentally alters the particle production mechanism—especially in the ultra-light regime—this effect does not yield a dominant dark matter component, at least, under the specific assumptions of our model.

Nevertheless, it is important to emphasize that these results establish a formal framework for understanding how global geometry scales the production of spectator fields. In this context, even a subleading abundance of light particles can have significant cosmological implications, as fluctuations in such spectator fields naturally translate into isocurvature perturbations. The high sensitivity of the spectrum to the spatial curvature at large scales, as found in our analysis, suggests that the presence of curvature could leave a distinct imprint on the isocurvature power spectrum. This is consistent with other studies exploring how global geometry modifies the low multipoles of the CMB~\mbox{\cite{Bonga2016,Bonga2017,Handley2019,Hergt2022}}, where the role of spatial curvature is known to be more prominent.

\section{Conclusions}
\label{sec:conclusions}

In this work, we have studied the phenomenon of cosmological particle production due to the expansion of spacetime in the early universe, in the context of FLRW geometries with spatial curvature different from zero, therefore extending previous studies that only considered flat spatial sections.

We considered a conformally coupled spectator scalar field  and analyzed its production during an inflationary period modeled by a de Sitter geometry, in order to obtain analytic results. We have considered two different prescriptions for implementing the transition from inflation to the subsequent radiation-dominated era: an instantaneous transition---reheating happens in one point in time---and a smooth transition implemented via the adiabatic vacuum prescription. In order to obtain the number density of produced particles during this process, we have chosen the Bunch-Davies vacuum as the initial state of the system, adapting the corresponding modes to the different coordinate charts that cover de Sitter spacetime, which correspond to FLRW geometries with closed, flat or open spatial sections. 

Our results demonstrate that spatial curvature has a tangible impact on particle production. Indeed, we find that production is enhanced in an open universe compared to the flat case, whereas it is suppressed in a closed universe. For light fields, these differences can reach several orders of magnitude in the associated relic abundance today, showing that the curvature of the spatial sections is a dominant factor in the infrared regime of the spectra

When restricting to curvature values consistent with current observational bounds, this effect is most prominent for ultra-light masses. While the resulting number density in this regime represents a subleading contribution to the total dark matter abundance, our findings underscore a fundamental sensitivity of the particle production mechanism to the global geometry of the universe. It is worth noting that even such subdominant abundances can have significant observational signatures; for instance, localized enhancements in the primordial spectra could trigger the formation of dense dark matter substructures, such as prompt cusps, potentially boosting indirect detection signals \cite{Olea-Romacho:2025mxj}.

Consequently, this work establishes that even minimal levels of spatial curvature, consistent with current observations, can fundamentally redefine the production efficiency of spectator fields. Such a signature is particularly relevant for the study of isocurvature perturbations, where the high sensitivity of the light-field spectra to spatial curvature at large scales could leave a distinct imprint on the CMB phenomenology. Our analytic framework thus provides a solid foundation for exploring more complex inflationary dynamics or non-minimal couplings, where the interplay between spatial curvature and early-universe particle production might reveal further non-trivial effects.

\section*{Acknowledgements}

This work is partially supported by the project PID2022-139841NB-I00 funded by MICIU/AEI/10.13039/50110001 1033 and by ERDF/EU, and by the R+D+I Project PID2023-149018NB-C44, funded by MICIU/AEI/10.13039/501100011033 and by ERDF/EU. This work is also part of the COST (European Cooperation in Science and Technology) Actions CA21106, CA21136, CA22113 and CA23130. Additionally, A.P.L. also acknowledges support through MICIU fellowship FPU20/05603, as well as from the Ramón Areces Foundation through its postdoctoral fellowships. 

\appendix

\section{Eigenfunctions of the Laplace-Beltrami operator}
\label{sec:eigenfunctions3Dcurved}

The eigenfunctions of the Laplace-Beltrami operator \eqref{eq:LaplaceBeltrami} are given  by  $\mathcal{H}_{klq}(\vec{u}) = \mathcal{U}_{kl}(u)Y_{lq}(\theta, \phi)$, where $\vec{u} = (u, \theta, \phi)$ and $Y_{lq}$ are the spherical harmonics in the unit 2-sphere, which are orthonormal and complete:
\begin{equation}
   \int \dd^2\Omega Y_{lq}(\theta, \phi)Y^*_{l'q'}(\theta, \phi)=\delta_{ll'}\delta_{qq'},
    \qquad
    \sum_{l=0}^{\infty}\sum_{q=-l}^l Y_{lq}(\theta, \phi) Y^*_{lq}(\theta^{\prime}\phi^{\prime}) = \delta(\cos{\theta} - \cos{\theta^{\prime}})\delta(\phi - \phi^{\prime}),
\end{equation}
The function  $\mathcal{U}_{kl}(u)$ reads
\begin{equation}
\mathcal{U}_{kl}(u)=\left\{
\begin{aligned} &\sqrt{\frac{M_{{\bar k}l}}{({\bar k}+1)^2\sin{\left(\sqrt{\abs{\kappa}}u\right)}}}P_{{\bar k}+1/2}^{-l-1/2}\left[\cos{\left(\sqrt{\abs{\kappa}}u\right)}\right]   &\quad \text{for} \quad \kappa>0, \\
 &\sqrt{\frac{2}{\pi}}j_l(ku) & \quad \text{for} \quad \kappa =0,\\
 &\sqrt{\frac{N_{{\bar k}l}}{{\bar k}^2\sinh{\left(\sqrt{\abs{\kappa}}u\right)}}}P_{i{\bar k} - 1/2}^{-l-1/2}\left[\cosh{\left(\sqrt{\abs{\kappa}}u\right)}\right] &\quad \text{for} \quad \kappa<0,
\end{aligned}
\right.
\label{eq:Eigenfunctions}
\end{equation}
with ${\bar k} = k/\sqrt{\abs{\kappa}}$.
The function $j_l $ is the spherical Bessel function of order $l$ 
and $P_{\nu}^{l}$ are the associated Legendre polynomials, where the order $\nu=i{\bar k} - 1/2$ corresponds to the conical or Mehler function \cite{Abramowitz1964}. At the same time, the normalization constants $M_{{\bar k}l}$ and $N_{{\bar k}l}$ read
\begin{equation}
M_{{\bar k}l}=\frac{({\bar k}+1)({\bar k}+l+1)!}{({\bar k}-l)!},\qquad
N_{{\bar k}l} = i^{-2l-1}{\bar k}^2\frac{\abs{\Gamma(i{\bar k}+l+1)}^2}{\abs{\Gamma(i{\bar k}+1)}^2},
\end{equation}
and are chosen such that we obtain the desired volume measure in momentum, as we will see shortly. Note that we follow a slightly different convention for the definition of the eigenfunctions~$\mathcal{H}_{klq}$ than the one used in \cite{Birrell1982}, more on the line of \cite{Abbott1986}, where the associated Legendre polynomials substitute the derivative of the cosine with respect to the hyperbolic cosine ($\kappa<0$), as it is derived in said reference. However, the range we consider here for the wavenumbers is that of~\mbox{\cite{Harrison1967, Parker1974}}, namely, those given in~\cref{eq:LabelsDescriptionA,eq:LabelsDescriptionB,eq:LabelsDescriptionC}.

As expected, all three eigenfunctions $\mathcal{U}_{kl}$ coincide when $\kappa \to 0$.  This can be straightforwardly checked noting that the asymptotic behavior of the Legendre polynomials for fixed $k$ in the limit $\kappa \to 0$ is
\begin{equation}
  P_{k/\sqrt{|\kappa|}+1/2}^{-l-1/2}\bigg[\cos \bigg(\sqrt{|\kappa|}u\bigg)\bigg] \sim \sqrt{\frac{2}{\pi}}  \bigg(k/\sqrt{|\kappa|}\bigg)^{-l} \bigg (\sqrt{|\kappa|}u\bigg)^{1/2} j_l\  (k  u )
\end{equation}
and similarly for complex order,
\begin{equation}
P_{ik/\sqrt{\abs{\kappa}}-1/2}^{-l-1/2}\left[\cosh{\left(\sqrt{\abs{\kappa}}u\right)}\right] \ {\underset{{\kappa \to 0}}{\simeq}} i^{l+1/2} \sqrt{\frac{2}{\pi}} \bigg(k/\sqrt{\abs{\kappa}}\bigg)^{-l} \bigg (\sqrt{|\kappa|}u\bigg)^{1/2} j_l(ku).
\end{equation}
Also, in this limit, for fixed $k$ and $l$,
\begin{equation}
    M_{{\bar k}l}\ {\underset{{\kappa \to 0}}{\simeq}} \bigg(k/\sqrt{\abs{\kappa}}\bigg)^{2l+2},\qquad
N_{{\bar k}l} = i^{-2l-1}\bigg(k/\sqrt{\abs{\kappa}}\bigg)^{2l+2}.
\end{equation}

The orthogonality of the eigenfunctions $\mathcal{H}_{klq}$ (in the generalized Dirac-delta sense for non-positive curvatures) is guaranteed by the orthonomality of the spherical harmonics and the following orthogonality expressions for the Bessel functions and the Legendre polynomials~\cite{Abramowitz1964,Harrison1967,Abbott1986,Ratra1995}:
\begin{align}
    &\int_{0}^{\infty}\text{d}u\,u^2 j_l(ku)j_l(k^{\prime}u)=\frac{\pi}{2k^2}\delta(k-k^{\prime}),
\label{eq:BesselOrthogonality}\\ &
\int_{-1}^1 \text{d}u P_{{\bar k}+1/2}^{-l-1/2}(u) P_{{\bar k}'+1/2}^{-l-1/2}(u) = 
\frac{({\bar k}-l)!}{({\bar k}+1)({\bar k}+l+1)!}\delta_{{\bar k}{\bar k}^{\prime}}.
\\
&  \int_{1}^{\infty}
P_{i{\bar k}-1/2}^{-l-1/2}(x)\,
P_{i{\bar k}'-1/2}^{-l-1/2,*}(x)\, dx
=
\frac{\pi}{{\bar k}\,\sinh(\pi {\bar k})\;
|\Gamma(l+1+i{\bar k})|^2}
\;\delta({\bar k}-{\bar k}').
\end{align}
Thus, one has that
\begin{equation}
\int\text{d}u\,r^2(u) \mathcal{U}_{kl}(u) \mathcal{U}_{k'l}(u)
=\begin{cases}
\displaystyle\abs{\kappa}^{-3/2}\frac{\delta_{{\bar k}{\bar k}^{\prime}}}{({\bar k}+1)^2}
, \quad &\text{for} \quad \kappa > 0\\ ~\\
\displaystyle\frac{\delta(k-k^{\prime})}{k^2}
, \quad &\text{for} \quad \kappa \leq 0
\end{cases}
\label{eq:Orthogonality}
\end{equation}

Additionally, the $\mathcal{U}_k(u)$ functions satisfy the completeness relations  
\begin{align} 
\int\diff\mu_k \,\mathcal{U}_{kl}(u) \mathcal{U}_{kl}(u')=\frac{\delta(u - u^{\prime})}{r(u)^2},
\end{align}
with $\displaystyle\int\diff\mu_k=\int \diff k\,k^2$ for $\kappa\leq 0$ and $\displaystyle\int \diff\mu_k=\sum_{{\bar k}=l}^\infty |\kappa|^{3/2}({\bar k}+1)^2$ for $\kappa>0$,
as can be easily checked from the completeness relations of the associated Bessel and Legendre functions and  
  shown in Ref. \citep{Bander1966} (making $u \to -iu$ and ${\bar k}+1 \to -i{\bar k}$ allows one to go from positive to negative spatial curvature). 
  This, together with the completeness of the spherical harmonics leads to the completeness of the eigenfunctions $\mathcal{H}_{klq}(u, \theta, \phi)$. For this we need Fubini's theorem for series \cite{SimonVol11972} that ensures that $\displaystyle\sum_{{\bar k}=0}^\infty\sum_{l=0}^{\bar k}=\sum_{l=0}^\infty\sum_{{\bar k}=l}^\infty$.

For our purposes, it is also useful to note that $\mathcal U_{kl}$ satisfies the addition rule \cite{Gradshteyn2014,Vilenkin1968} (which we have  verified numerically as well)
\begin{equation}
    \sum_{l=0}^K (2l+1)\mathcal{U}_{kl}^2(u)=\frac2\pi,
    \label{eq:sumruleU}
\end{equation}
where $K=\infty $ for $\kappa\leq 0$, $K=\bar k$ for positive $\kappa$.


\bibliographystyle{JHEP.bst}
\bibliography{references.bib}

\end{document}